\documentclass[aps,prl,twocolumn,showpacs]{revtex4}
\usepackage{epsfig}
\usepackage{graphicx}

\begin{document}

\DeclareGraphicsExtensions{.eps,.EPS}

\title{All-Optical Production of Chromium Bose-Einstein Condensates}
\author{Q. Beaufils, R. Chicireanu, T. Zanon, B. Laburthe-Tolra, E. Mar\'echal, L. Vernac, J.-C. Keller, and O. Gorceix}
\affiliation{Laboratoire de Physique des Lasers, UMR 7538 CNRS,
Universit\'e Paris Nord, 99 Avenue J.-B. Cl\'ement, 93430
Villetaneuse, France}

\begin{abstract}
We report on the production of $^{52}$Cr Bose Einstein Condensates
(BEC) with an all-optical method. We first load 5.10$^6$ metastable
chromium atoms in a 1D far-off-resonance optical trap (FORT) from a
Magneto Optical Trap (MOT), by combining the use of Radio Frequency
(RF) frequency sweeps and depumping towards the $^5S_2$ state. The
atoms are then pumped to the absolute ground state, and transferred
into a crossed FORT in which they are evaporated. The fast loading
of the 1D FORT (35 ms 1/e time), and the use of relatively fast
evaporative ramps allow us to obtain in 20 s about 15000 atoms in an
almost pure condensate.
\end{abstract}

\date{\today}
\maketitle

The study of the degenerate quantum phases of chromium is especially
appealing for two main reasons. First, the atomic magnetic moment of
6 $\mu_B$ (Bohr magneton) leads to large anisotropic long range
dipole-dipole interactions, which are non negligible compared to the
contact interaction \cite{Dip Dip Interaction}, and can even become
the dominant interaction close to a Feshbach resonance \cite{Strong
Dip Int}. In this regime, the stability and excitation properties of
dipolar BECs are completely modified by dipole-dipole interactions
\cite{Dip stability}. In addition, the large $S=3$ spin in the
ground state makes Cr a unique element for spinor physics
\cite{Spinor}. Second, the existence of a fermionic isotope
($^{53}$Cr, 10 \% natural abundance) opens the way to obtain a
degenerate dipolar Fermi sea, and to study the interesting stability
properties of a dipolar boson-fermion mixture
\cite{FeStabilization}.

The historic \cite{BEC 87Rb} and still conventional way to produce
quantum degenerate gases is evaporation inside a magnetic trap (MT).
An other possibility, demonstrated first for Rb \cite{Barrett}, is
to evaporatively cool in an optical trap created by a far red
detuned laser. These traps offer an interesting experimental
alternative as the highly confining MTs required to evaporate
efficiently demand either large currents, or the use of integrated
structures \cite{BEC chips}. For some atoms, the winning strategy to
obtain condensation has been to use a FORT, either because of high
inelastic collision rates (for Cs \cite{In BEC Cs} and Cr \cite{In
Coll Cr},\cite{BEC Cr}), or because of the absence of a permanent
magnetic moment (for Yb \cite{BEC Yb}). In the first case optically
pumping the atoms to the lowest energy Zeeman substate suppresses
all two-body inelastic collisions at low temperature, but these high
field seeking states cannot be trapped magnetically: the use of
optical traps is necessary. The evaporation is then performed in a
crossed FORT with a standard procedure, for which the evaporation
dynamics is well understood \cite{EvCoolOptical}.

However, efficiently loading a FORT is not straightforward in
general and especially for Cr. In particular in our experiment, a
direct loading of a Cr optical trap in the ground $^7S_3$ state from
a MOT leads to small number of atoms, presumably because of a high
light assisted inelastic collision rate \cite{MOTCr2,PRA1}. The
loading procedure used to obtain the first Cr BEC \cite{BEC Cr} was
to accumulate the atoms in metastable D states inside a MT, before
transferring them first into an elongated Ioffe Pritchard MT, and
then in a 1D FORT (produced by one beam). Our strategy is quite
different, as we directly load a 1D FORT of metastable atoms from
our MOT. In this article we first describe our original method to
rapidly load a 1D FORT (in less than 100 ms, to be compared to about
20 s in \cite{BEC Cr}). We then describe the evaporation procedure
and show the evidence for the production of a BEC.

We produce an atomic Cr beam from an oven running at
1500${^\circ}$C. After a one meter long Zeeman Slower atoms are
captured in a standard MOT, with a few 10$^6$ atoms and an initial
phase space density of 5.10$^{-7}$. An horizontal retroreflected 35
W IR laser beam (produced by a 50 W fiber laser, at 1075 nm) is
focused at the MOT center (with a waist of 42 $\mu$m). The 425.5 nm
cooling transition from the ground $^7S_3$ to the excited $^7P_4$
state has leaks towards metastable $D$ states. When atoms decay into
these $D$ states in the low field seeking substates, they remain
trapped due to the strong confinement of the IR laser along two
transverse directions, and to the magnetic gradient created by the
MOT coils along the IR beam propagation axis. We reported in
\cite{art EPJD} how we could obtain about one million metastable
atoms in this continuous FORT loading procedure. However
encouraging, this result turned out to be insufficient to reach
quantum degeneracy: we obtained a final phase space density in the
5.10$^{-4}$ range after evaporation in a crossed FORT. Nevertheless
as the dependance of the evaporation process with the initial number
of atoms is highly non linear, we estimated using
\cite{EvCoolOptical} that a gain of a factor about 5 in the 1D FORT
loading could be enough to reach degeneracy.

As a first main progress we were able to lower the magnetic forces
applied to the atoms during the loading in order to 1) allow the
capture of metastable atoms produced in any magnetic substate, and
2) increase the volume of the trap. We report in \cite{rampes RF}
how we use fast RF frequency ramps to spin flip the atoms in the
trap at a high rate: we excite the atoms with a RF produced by a 150
W amplifier that we send to a 8-turn, 8 cm diameter coil located 5
cm from the MOT center, and we continuously sweep the RF frequency
while loading the FORT. Magnetic forces are thus averaged out, and
we can trap in a pure 1D FORT 80$\%$ more atoms than without RF.

In a second decisive step, we investigated the possibility to
accumulate atoms in a different metastable state. The $^5S_2$ state
is easier to load due to the larger branching ratio to this level
from $^7P_3$ than to the $D$ states from $^7P_4$ (a gain of about
100 is expected \cite{NIST}). Besides, our study in \cite{art EPJD}
proved that one of the main limiting processes for the continuous
loading in the FORT in metastable $D$ states comes from inelastic
collisions, and the absence of spin-orbit interaction at first order
for $^5S_2$ is thus favorable. Finally, calculations of the light
shifts show that optical trap depths are expected to be almost twice
larger in the $^5S_2$ state than in the D states.

To depump towards $^5S_2$, we apply during the loading process a
weak beam at 427.6 nm (35 $\mu$W, 3 mm 1/e$^2$ diameter), at
resonance with the $^7S_3\rightarrow^7P_3$ transition. In order to
guarantee that the MOT capture efficiency is not reduced by this new
beam, and to repump any atom in the $^5S_2$ state outside of the
FORT, we add a "dark spot" repumper at 633 nm: we shine on the atoms
a 0.4 mW beam going across an horizontal wire imaged on the MOT
center, with an image size of 400 $\mu$m. The depumping effect of
the 427.6 nm beam is thus counterbalanced in most of the MOT capture
volume by the 633 nm beam (the MOT beams have a 1/e$^2$ diameter of
7 mm), but the atoms which accumulate in the 1D optical trap in the
$^5S_2$ state do not get repumped back to the ground state. The use
of this dark spot during the loading increases the final number of
atoms by 20 $\%$. A similar technique is used in \cite{Lett} to
reduce the amount of inelastic collisions in a Na cloud.

In order to characterize the 1D FORT loading, and for evaporation,
we then prepare the atomic sample in the $^7S_3$, $m_J$=-3 absolute
ground state. We switch off the MOT beams and magnetic gradients, as
well as the RF, and we repump the atoms in $^5S_2$ and $^5D_4$ to
the ground state. We do not have yet the possibility to repump atoms
in the $^5D_2$ and $^5D_3$ states, which may cause some inelastic
collisions during evaporation. Finally, the atoms are transferred to
$m_J$=-3: after the 20 ms necessary for the eddy currents generated
by switching off the MOT coils to disappear, we apply with the 427.6
nm laser a 50 $\mu$s, 0.25 mW retro-reflected pulse of circular
polarization, in presence of a 2.3 G magnetic field. The field
direction during optical pumping is finely aligned with the
polarization beam propagation axis, so that we can send ms long
pulses without losing much atoms, proving that indeed a dark state
has been obtained on this optical $J\rightarrow J$ transition. We
image atoms in this state using a dark ground absorption imaging
system with a resonant circularly polarized beam on the trapping
transition.

Results shown in Fig. 1 have been obtained by first forming a
stationary MOT (in about 50 ms) in presence of the depumping and
dark spot beams, and then switching on the horizontal IR beam for a
given duration. The number of atoms trapped typically reaches 2.5
millions when no RF ramps are applied, and about 4.5 millions with
RF ramps. In this case the 1/e loading time is about 35 ms, and from
the slope at $t=$0, we infer a loading rate equal to 1.3 10$^8$
atoms.s$^{-1}$, which is only a factor about 4 smaller than the MOT
one. We have thus realized a very efficient loading procedure from a
MOT into a 1D FORT, with a final temperature of 100 $\mu$K, a phase
space density of 5.10$^{-6}$, and an elastic collision rate of 50
s$^{-1}$. Compared to typical starting parameters for alkalis atoms
in FORTs before evaporation (see for example \cite{Barrett}), this
phase space density is relatively small (presumably due to large
inelastic losses), which will lead to substantially longer
evaporation durations.

\begin{figure}[h]
\centering
\includegraphics[width=3.5in]{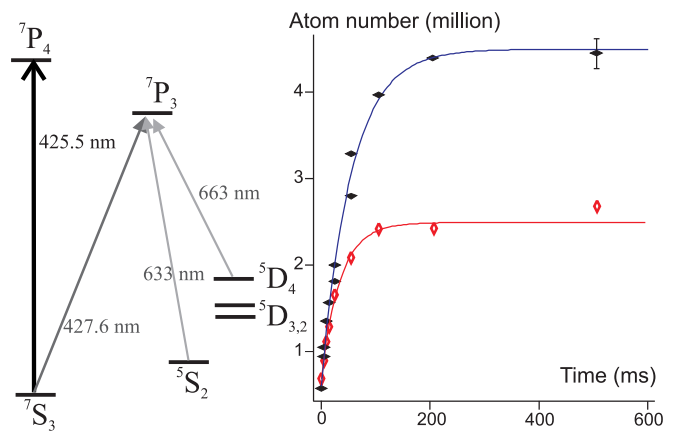}
\caption{\setlength{\baselineskip}{6pt} {\protect\scriptsize Loading
of the 1D horizontal FORT (right). We plot the number of metastable
atoms after a given accumulation duration. The presence of RF
frequency ramps during the accumulation almost doubles the final
number of atoms (top curve versus bottom). The solid lines are
exponential fits. The error bar gives (as in the other figures) the
typical statistical uncertainty. We show (left) the atomic levels
and transitions of interest.}} \label{1DLoading}
\end{figure}

\begin{figure}[h]
\centering
\includegraphics[width=3.5in]{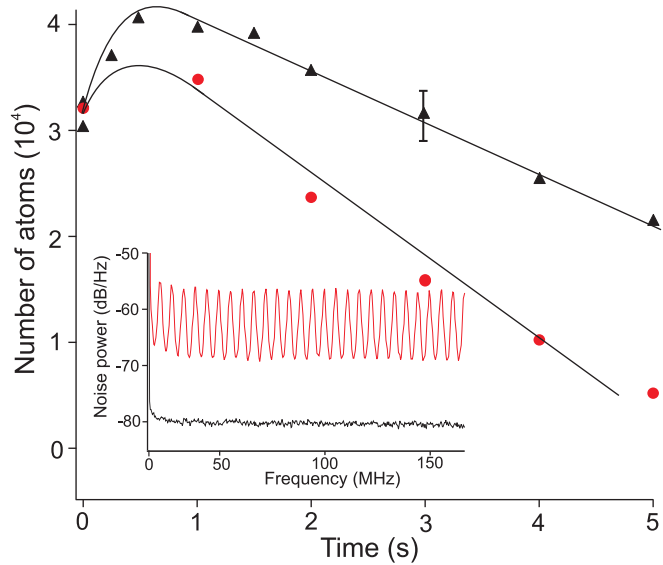}
\caption{\setlength{\baselineskip}{6pt} {\protect\scriptsize Top:
Experimental evidence for the importance of the relative
polarizations of the different IR beams creating the crossed FORT.
After a dimple has been formed by transferring half of the IR power
to the vertical beam, the number of atoms in the dimple is plotted
versus time. When the 3 IR beams (the 2 horizontal ones, and the
vertical) have a linear polarization along 3 orthogonal axis
(triangles), the loading from the horizontal FORT is more efficient
than when two have parallel polarizations (circles). Besides, the
losses are reduced. Solid lines are guides for the eye. Bottom:
Intensity noise spectrum of the IR laser observed with a 1 GHz
bandwidth photo detector (top curve); the noise spectrum obtained
with no light is shown below.}} \label{PolarEffect}
\end{figure}

To reach degeneracy, we start to form a crossed FORT just after
preparing the atoms in the $^7S_3$, $m_J$=-3 state, thus creating a
"dimple" \cite{dimple}: we transfer some laser power from the
horizontal beams to a second IR beam, which is almost vertical, and
has a 56 $\mu$m waist. The power dispatching is controlled via a
half wave plate on a computer controlled rotation stage, in front of
a polarizing beam splitter cube: we proceed to a 9.1 s long linear
rotation of this plate by an angle equal to 32${^\circ}$. We stress
that the relative polarization of the three IR beams does matter.
The dimple loading efficiency is optimal when the polarizations of
the 3 IR beams are orthogonal to better than about $20{^\circ}$.
When the vertical beam polarization is parallel to one of the
horizontal beams polarizations, we observe a strong reduction of the
loading efficiency of the crossed trap (see Fig. \ref{PolarEffect})
which prevents to reach BEC \cite{note}. Yet the path differences
between the IR beams are much larger than the expected coherence
length of our IR laser, assuming a 5 nm full-width-half-max
continuous emitting spectrum. Our observations may be related to the
intensity noise spectrum of the IR laser (see insert in Fig.
\ref{PolarEffect}), which points towards a regular structure in its
frequency spectrum.

\begin{figure}[h]
\centering
\includegraphics[width=3.5in]{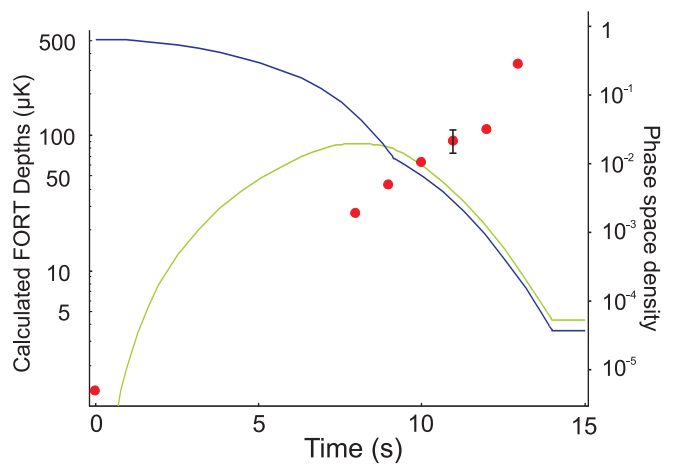}
\caption{\setlength{\baselineskip}{6pt} {\protect\scriptsize
Calculated potential depths created by the horizontal and vertical
(bell shape) FORT beams (Log scale) during the transfer in the
crossed FORT and the evaporation. We show as well the classical
phase space density (dots) in the initial 1D FORT (at t=0), and
after all the atoms are transferred into the dimple. At the very end
of the ramp, the cloud reaches degeneracy, and the classical phase
space density (exceeding 1) is therefore not shown.}} \label{Rampes}
\end{figure}

During the dimple formation, we start to lower the total IR laser
power using an acousto-optic modulator (AOM): 6 s after the atomic
polarization, a decreasing non linear IR laser power ramp is
triggered. This ramp corresponds to a linear ramp of the voltage
sent to the power input of the voltage controlled oscillator driving
the AOM. The resulting theoretical depths of the traps created by
the horizontal and vertical FORT beams are shown in Fig.
\ref{Rampes}. We show as well the evolution of the experimental
classical phase space density
$N_{at}(\frac{\hbar\overline{\omega}}{k_{B}T})^3$, where
$\overline{\omega}$ is the geometric means of the dimple trap
angular frequencies, and $T$ the temperature of the cloud containing
$N_{at}$ atoms.

The parameters characterizing the magnetic field around the trap
center turned out to be critical to obtain a BEC. We need to
compensate for a magnetic potential curvature along the weak axis of
the 1D FORT (Ox) due to the coil producing the polarization field,
by using a 4 cm radius coil located 10 cm above the MOT center, in
an horizontal plane. The 1D FORT life time is thus significantly
increased. This new coil creates as well a gradient b$_z$ along the
vertical axis, which induces a force opposed to gravity. We do not
obtain the most efficient evaporation when the two forces compensate
as expected according to \cite{Comparat}, but for 6$\mu_B$b$_z$=-0.8
$m_{52}$ $g$, where $m_{52}$ is the atom mass, and $g$ is the
gravity acceleration. This effect can be interpreted by the fact
that it is better if the atoms evaporated out of the dimple into the
wings of the crossed FORT do not come back to collide with atoms
remaining in the dimple \cite{Gravity helps}. A leak along these
wings is thus favorable, and we found indeed that it is better to
have as well a small gradient along Ox (0.1 G.cm$^{-1}$) once the
dimple is formed.

\begin{figure}[h]
\centering
\includegraphics[width=3.5in]{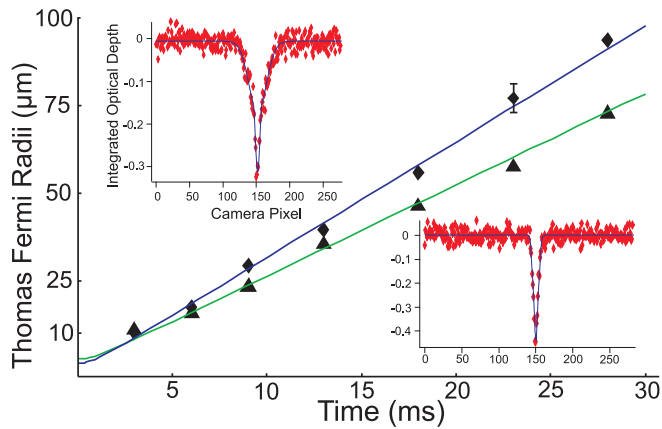}
\caption{\setlength{\baselineskip}{6pt} {\protect\scriptsize
Analysis of the experimental free expansion of an almost pure BEC.
The Thomas Fermi radii along y (diamonds) and z (triangles) are
deduced from absorption pictures. The solid lines are results of
numerical simulations using the measured trap frequencies, with the
in situ radii giving the best agreements with the experimental data.
Inserts: integrated optical density of the atomic cloud after 4 ms
of free expansion, when $T_c$ is reached (left), and well below
$T_c$ (right); one pixel corresponds to 2.2 $\mu$m.}}
\label{ExpAnis}
\end{figure}

Phase transition to BEC is observed with the appearance of a very
narrow feature at the center of the velocity distribution, as
measured by the cloud's profile after some free expansion. A bimodal
velocity distribution (see the left insert of Fig. \ref{ExpAnis}) is
observed below $T=150$ nK, consistent with a predicted degeneracy
temperature of 170 nK \cite{Stringari}. We estimate the trap
frequencies at the critical point by scaling the experimental
frequencies ($f_1$=$f_2=110\pm2$ Hz, $f_3=150\pm2.5$ Hz) measured
through parametric excitation at the very end of the evaporation
ramp, where we obtain an almost pure condensate of up to
$N_{BEC}$=15000 atoms (right insert of Fig. \ref{ExpAnis}).

In order to characterize the pure BEC we analyzed the absorption
pictures along the two axis (Oy and Oz) of the CCD camera during
free expansions as long as 30 ms. We obtain the two radii R$_{y}$(t)
and R$_{z}$(t) by standard analysis, assuming the Thomas Fermi (TF)
regime \cite{Stringari}. The corresponding results in Fig.
\ref{ExpAnis} demonstrate a clear anisotropic expansion. A
comparison with a numerical resolution of the Gross-Pitaevskii
equation \cite{CastinDum} provides the following in situ TF radii:
R$_{zTF}= 5\pm0.25$ $\mu$m (=R$_{xTF}$), R$_{yTF}=3.75\pm0.25$
$\mu$m, and the chemical potential
$\mu/h=m_{52}\overline{\omega}^2\overline{R}_{TF}^2/2=790\pm90$ Hz
($>>f_3$, which justifies the TF approximation), $\overline{R}_{TF}$
being the geometric means of the TF radii. The value of $\mu$ is
consistent with the expected chemical potential (
$\mu_{TF}/h=865\pm40$ Hz), given $N_{BEC}$, $\overline{\omega}$, and
the value of the scattering length reported in \cite{a52Cr}. The
peak density in the BEC is $(6.4\pm0.75).10^{13}$ cm$^{-3}$.

We want to emphasize the relative simplicity of our setup and of the
experimental procedure we use to reach BEC with chromium. First, the
total duty cycle for producing a BEC is about 20 s (while it is more
than 35 s in \cite{BEC Cr}). It could be as small as 14 s since the
1D FORT loading takes about 100 ms, but we need some time at the end
of a cycle to recover the starting parameters. Thanks to the
relatively short evaporation duration, vacuum requirements to reach
degeneracy are easier to fulfill than in many BEC experiments: the
pressure in our experimental chamber is in the 5.10$^{-11}$ mbar
range, which induces a limited 1/e life time of 25 s for the 1D
FORT. In addition, we run our oven at a temperature limited to
1500${^\circ}$C (1600${^\circ}$C in \cite{BEC Cr}), increasing thus
both the robustness of our setup, and its capacity to deliver a Cr
beam over a long period of time.

In conclusion, we have obtained a Cr BEC with an original strategy.
The key point was to load from a MOT a 1D FORT with a sufficient
atom number, which could not be obtained by running the oven at a
higher temperature. Reaching degeneracy thus required the
development of two new accumulation techniques described in this
paper. They could also be instrumental for achieving BEC with other
atoms, such as calcium, which can be optically trapped in metastable
states \cite{Ca}. Besides, we point out that the non perfect
compensation of gravity, and the orthogonality of the polarizations
of the IR beams involved in the crossed FORT, are necessary for a
successful evaporation in optical traps.

\vspace{1cm}

Acknowledgements: LPL is Unit\'e Mixte (UMR 7538) of CNRS and
of Universit\'e Paris Nord. We acknowledge financial support from Conseil R\'{e}%
gional d'Ile-de-France, Minist\`{e}re de l'Education, de
l'Enseignement Sup\'{e}rieur et de la Recherche, European Union and
IFRAF. We thank R. Barb\'{e} and A. Pouderous for their
contributions to the experiment. We acknowledge technical assistance
from the mechanical shop (G. Baqu\'{e}, M. Fosse, A. Kaladjian), and
the electronic shop (F. Wiotte, J. de Lapeyre).


\begin{thebibliography}{99}
\bibitem{note} We also measured a heating rate in the $\mu$K.s$^{-1}$
range for a crossed FORT consisting of beams crossing in an
horizontal plane with non orthogonal polarizations. Similar effects
have been observed by other groups, but have not been reported to
our knowledge.
\end{thebibliography}
\end{document}